\documentclass[reprint,superscriptaddress,aps,prd,floatfix,nofootinbib,bibnotes]{revtex4-2}
\usepackage[colorlinks=true,linkcolor=black,citecolor=blue, urlcolor=blue,bookmarks=false]{hyperref}
\hypersetup{breaklinks=true}
\usepackage[utf8]{inputenc}
\usepackage{natbib,slashed}
\usepackage{bm}
\usepackage{dcolumn}
\usepackage{array}
\usepackage{amsmath,mathtools}
\usepackage{amssymb}
\usepackage{multirow}
\usepackage{graphicx,subcaption}
\usepackage{tensor}
\usepackage{comment}
\usepackage{enumitem}
\graphicspath{{./fig/}}
\usepackage{float}
\usepackage[normalem]{ulem}  
\usepackage{color} 

\renewcommand{\sout}{\bgroup \color{red} \ULdepth=-.5ex \ULset}

\begin{document}
\title{The Inevitable Quark Three-Body Force and its Implications for Exotic States}

\author{Sungsik Noh}
\email{sungsiknoh@kangwon.ac.kr}
\affiliation{Division of Science Education, Kangwon National University, Chuncheon 24341, Korea}

\author{Aaron Park}
\email{aaron.park@yonsei.ac.kr}
\affiliation{Department of Physics and Institute of Physics and Applied Physics, Yonsei University, Seoul 03722, Korea}

\author{Hyeongock Yun}
\email{mero0819@yonsei.ac.kr}
\affiliation{Department of Physics and Institute of Physics and Applied Physics, Yonsei University, Seoul 03722, Korea}

\author{Sungtae Cho}\email{sungtae.cho@kangwon.ac.kr}
\affiliation{Division of Science Education, Kangwon National University, Chuncheon 24341, Korea}

\author{Su Houng Lee}%
\email{suhoung@yonsei.ac.kr}
\affiliation{Department of Physics and Institute of Physics and Applied Physics, Yonsei University, Seoul 03722, Korea}

\date{\today}
\begin{abstract}

Three-body nuclear forces are essential for explaining the properties of light nuclei with a nucleon number greater than three. Building on insights from nuclear physics, we extract the form of quark three-body interactions and demonstrate that these terms are crucial for extending the quark model fit of the meson spectrum to include baryons using the same parameter set.  We then discuss the implications of our findings for exotic configurations involving more than three quarks, such as the $T_{cc}$ and $\chi_{c1}(3872)$.  We find that the quark three-body interactions provide additional repulsion on the order of 10 MeV for the compact configurations of both the $T_{cc}$ and $\chi_{c1}(3872)$. This result, combined with previous calculations, strongly suggests that these tetraquark states are molecular rather than compact states. 
\end{abstract}

\maketitle

{\it Introduction:}
It has long been known that nuclear three-body forces exist\cite{Primakoff:1939zz}, and their presence is crucial in correctly reproducing the binding energy of few-nucleon systems\cite{Carlson:1983kq}. Three-nucleon forces also arise naturally in chiral effective field theory\cite{Epelbaum:2005pn}. The first calculation of the three-nucleon force was attributed to the two-pion exchange among three nucleons \cite{Fujita:1957zz}. In particular, the model where the intermediate nucleon becomes an excited nucleon, such as the delta, has been established since the 1960s\cite {Brown:1968qla, Brown:1969xqx,Yang:1974zz}. 

\begin{figure}[h]
\centering
\includegraphics[scale=0.8]{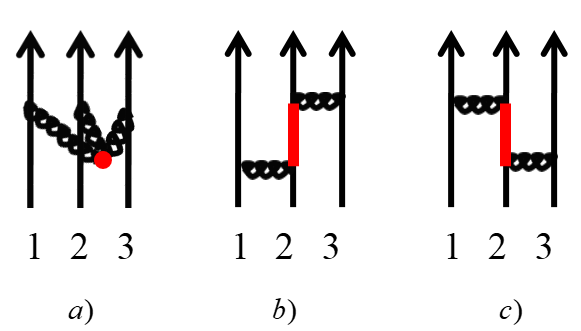}
\caption{Quark three-body interaction from a) three-gluon interaction and b), c) two-gluon exchange with excited quark intermediate state.} 
\label{Three-body}
\end{figure}

The three-body forces among quarks have also been discussed for some time\cite{Desplanques:1992rf}. These are the three-color matrix type proportional to the two structure constants of color SU(3), namely $f^{abc}$ and $d^{abc}$\cite{Dmitrasinovic:2001nu}.
The first type $f^{abc}\lambda^a_1 \lambda^b_2 \lambda^c_3$, shown in Fig.\ref{Three-body}-a), where the subscripts denote the three quarks, arises naturally due to the three-gluon interaction.  On the other hand, this interaction is not SU(3) invariant and does not contribute to the baryon mass.

The $d$-type does contribute to the baryon mass, 
with its strength constrained by the stability of the baryon mass.
The effects of this interaction on the nucleon mass and on the stability of exotic states have been discussed in previous studies \cite{Pepin:2001is,Park:2015nha}.  
However, this term contributes uniformly to all the flavor octet baryon states in the color singlet representation and hence does not improve the fit to the flavor-dependent masses. Furthermore, the origin of this term has not been discussed, and the rationale behind introducing negative signs for interactions involving antiquarks remains unclear\cite{Dmitrasinovic:2001nu}. 

At the same time, it is known that the quark masses required to describe the baryon spectrum differ from those used in the meson spectrum in a simple quark model\cite{Karliner:2014gca}.  Also, the naive extrapolation of the color-spin interaction in the meson and baryon systems does not seem to follow the necessary changes in the color-spin matrix elements\cite{Lee:2007tn}.  These observations imply that additional effects need to be considered when transitioning from a two-quark to a three-quark system. The inclusion of a quark three-body interaction naturally emerges as a necessary addition.

In this work, we will draw upon the wisdom from nuclear physics, where the three-body interaction originates from two separate pion exchanges with an intermediate delta\cite{Yang:1974zz}. Here, the pions and nucleons are replaced by effective gluon exchange and quark diagrams, as shown in Fig.\ref{Three-body} b) and c).
As we will see, the newly introduced quark three-body interactions exhibit additional spin dependencies and are found to be indispensable in consistently describing the meson and baryon spectrum on the same footing. These interactions are crucial for understanding the stability of multiquark states.

{\it Quark model Hamiltonian:} The quark model Hamiltonian typically has the following form\cite{Park:2015nha,Park:2016cmg,Park:2016mez,Park:2017jbn,Park:2018wjk,Park:2019bsz,Park:2021hqb,Noh:2021lqs,Noh:2023zoq}. 
\begin{eqnarray}
H &=& \sum^{n}_{i=1} \left( m_i+\frac{{\mathbf p}^{2}_i}{2 m_i} \right)-\frac{3}{4}\sum^{n}_{i<j}\frac{\lambda^{c}_{i}}{2} \,\, \frac{\lambda^{c}_{j}}{2} \left( V^{C}_{ij} + V^{CS}_{ij} \right),
\label{Hamiltonian-two}
\end{eqnarray}
where $n$ is the number of quarks in the hadron and $m_i$ is the mass of the $i$ quark. The two-body quark potential is composed of color-color interaction  $V^C_{ij}(r_{ij})$ and color-spin interaction that has an additional spin dependence written as $V^{CS}_{ij}=\frac{\sigma_i \cdot \sigma_j}{m_i m_j}V^S(r_{ij})$ typical of a magnetic type.  

Let us now consider the quark three-body interaction. It can be obtained by utilizing the two-body interactions provided in Eq.\eqref{Hamiltonian-two} at the vertices for the diagrams depicted in Fig.\ref{Three-body} b) and c). 
After adding the contribution where all three quarks (1,2,3) are the intermediate quarks
and using the properties of the color and spin matrices, we obtain the following interactions.  
\begin{widetext}
\begin{eqnarray}
L^{C-C}_{123} &= &  \frac{4}{3} \bigg(
 \frac{\lambda_2^c  \lambda_3^c}{m_1}
+ \frac{\lambda_1^c  \lambda_3^c}{m_2} 
+\frac{\lambda_1^c  \lambda_2^c}{m_3} \bigg) + 2 d^{abc}(\lambda_1^a  \lambda_2^b   \lambda_3^c) \bigg(
 \frac{1}{m_1}
+ \frac{1}{m_2} 
+\frac{1}{m_3} \bigg),
\nonumber \\
 L^{S-S}_{123} &= & \frac{1}{m_1m_2m_3} \bigg[  \frac{4}{3} \bigg( \frac{(\sigma_2 \cdot \sigma_3)(\lambda_2^c \lambda_3^c)}{m_1^2} 
+ \frac{(\sigma_1 \cdot \sigma_3)(\lambda_1^c \lambda_3^c)}{m_2^2}
+\frac{(\sigma_1 \cdot \sigma_2)(\lambda_1^c \lambda_2^c)}{m_3^2} \bigg) 
\nonumber \\
&&+
2  d^{abc}(\lambda_1^a  \lambda_2^b   \lambda_3^c)  \bigg( \frac{\sigma_2 \cdot \sigma_3  }{m_1^2} +\frac{\sigma_1 \cdot \sigma_3 }{m_2^2}+\frac{\sigma_1 \cdot \sigma_2 }{m_3^2} \bigg) 
 -2 \epsilon_{ijk} \sigma^i_1 \sigma^j_2 \sigma^k_3 f^{abc} \lambda^a_1 \lambda^b_2 \lambda^c_3 \bigg( \frac{1}{m_1^2} +\frac{1}{m_2^2}+\frac{1}{m_3^2} \bigg) \bigg],
\nonumber \\
L^{C-S}_{123} &= &  \frac{4}{3}  \bigg[    \frac{(\lambda_2^c \lambda_3^c)}{m_1} \bigg( \frac{\sigma_1 \cdot \sigma_3}{m_1m_3} + \frac{\sigma_1 \cdot \sigma_2}{m_1m_2}  \bigg) +\frac{(\lambda_1^c \lambda_3^c) }{m_2} \bigg( \frac{\sigma_2 \cdot \sigma_3}{m_2m_3} + \frac{\sigma_1 \cdot \sigma_2}{m_1m_2}  \bigg) 
+    \frac{(\lambda_1^c \lambda_2^c)}{m_3} \bigg( \frac{\sigma_2 \cdot \sigma_3}{m_2m_3} + \frac{\sigma_1 \cdot \sigma_3}{m_1m_3}  \bigg) \bigg] 
\nonumber \\
& & +     2d_{abc} (\lambda_1^a
 \lambda_2^b \lambda_3^b )  \bigg[   \frac{1}{m_1} \bigg( \frac{\sigma_1 \cdot \sigma_3}{m_1m_3} + \frac{\sigma_1 \cdot \sigma_2}{m_1m_2}  \bigg) +\frac{1 }{m_2} \bigg( \frac{\sigma_2 \cdot \sigma_3}{m_2m_3} + \frac{\sigma_1 \cdot \sigma_2}{m_1m_2}  \bigg) 
+    \frac{1}{m_3} \bigg( \frac{\sigma_2 \cdot \sigma_3}{m_2m_3} + \frac{\sigma_1 \cdot \sigma_3}{m_1m_3}  \bigg) \bigg] .
\label{eq:l3}
\end{eqnarray}
\end{widetext}
Here, the superscripts $C$ and $S$ show the two-body color-color and color-spin interactions, respectively, used at the vertices.
As one can see, there are spin-dependent contributions to the quark three-body forces that have not been considered before.

Now, because of the anti-commutation relation between the color matrix $\bar{\lambda}$ of an antiquark, the term proportional to $d^{abc}$ in Eq.~\eqref{eq:l3} will come with a negative $(-)$ sign if the intermediate state is an antiquark.  For example, in the first line of Eq.~\eqref{eq:l3}, if particle 2 is an antiquark, the second term will change to $2 d^{abc}(\lambda_1^a  \bar{\lambda}_2^b   \lambda_3^c) \bigg(
 \frac{1}{m_1}
- \frac{1}{m_2} 
+\frac{1}{m_3} \bigg)$.  Similar changes also apply to the second term in $L^{S-S}$ and $L^{C-S}$.   Such a rule automatically leads to a charge conjugation symmetric three-body interaction, which was previously enforced artificially\cite{Dmitrasinovic:2001nu}.

The three-quark color interaction can be calculated using permutation matrices when all the particles are quarks. For the diagonal component of a multiquark state, it is given as follows\cite{Park:2022jpf}:.
\begin{align}
    \sum_{i<j<k}d^{abc}\lambda^a_i \lambda^b_j \lambda^c_k =& \frac{4}{3}C^{(3)} -\frac{10}{3}C^{(2)}+\frac{80}{27}n, \label{d-type}\\
    \sum_{i<j<k}f^{abc}\lambda^a_i \lambda^b_j \lambda^c_k =& 0, \label{f-type}
\end{align}
where $C^{(3)}=\frac{1}{18}(p-q)(2p+q+3)(p+2q+3) $ and $C^{(2)}=\frac{1}{3}(p^2+q^2+pq+3p+3q)$ are the cubic and quadratic Casimir invariants of SU(3).
When there are antiquarks involved, one should return to the (anti-)commutation of two color matrices to obtain the matrix elements instead of using Eq.~\eqref{d-type} and~\eqref{f-type}. 

For a color singlet state, the Casimir invariants vanish. Thus, Eq.~\eqref{d-type} will be proportional to the total number of quarks, $n$, multiplied by a universal constant common for a color singlet state $ d^{abc}(\lambda_1^a  \lambda_2^b   \lambda_3^c) = \frac{80}{9}$. In such cases, as with the two-body color-color interaction, the multiquark configuration does not gain any attraction compared to the configuration where it decomposes into two color singlet states with quark numbers $n_1$ and $n-n_1$, respectively.  That is why the extra spin dependencies are important, as they contribute to possible additional attraction when combined into a compact multiquark configuration.  
Such non-linearity in the number of quarks also occurs for the two-body color-spin interaction, whose matrix elements for color singlet multiquark states include terms proportional to  $n^2$\cite{Aerts:1977rw}.

Let us now consider the effects of the three-quark interactions on the color singlet baryons.  In this case,
we take all the quark masses to be equal to $m_q$, use $\lambda_i^c  \lambda_j^c = -\frac{8}{3}$, $f^{abc} \lambda_1^a \lambda_2^b \lambda_3^c=0$, and note  $\sigma_1 \cdot  \sigma_2 +\sigma_2 \cdot  \sigma_3 + \sigma_1 \cdot  \sigma_3$ is equal to $+3$ and $-3$ for the $\Delta$  and nucleon, respectively.  Assuming the overall strength multiplying the interactions in Eq.~\eqref{eq:l3} are $A,B$ and $C$, respectively, we have
\begin{eqnarray}
H_{N,\Delta} &= &  A \frac{128}{3m_q}   \mp B \frac{128}{3m_q^5} 
 \mp   C \frac{256}{3m_q^3} ,  \label{eq:pd}
\end{eqnarray}
where $\mp$ corresponds to the nucleon and Delta, respectively. One notes that the spin-dependent three-quark interactions are important for better fitting the nucleon-delta mass difference, as they induce mass splitting. Previous works assumed that the quark three-body interaction was not important because, when only the color-color type is considered, only the overall mass shift is affected through the term proportional to 
$A$ in Eq.~\eqref{eq:pd}\cite{Capstick:1986ter}.

{\it Meson spectrum:} To see why such quark three-body interactions are essential to consistently describe the baryon spectrum, 
consider first using only the two-quark interaction terms in the Hamiltonian given in Eq. \eqref{Hamiltonian-two} with the following forms that were extensively used before. 
\begin{eqnarray}
V^{C}_{ij} &=& - \frac{\kappa}{r_{ij}} + \frac{r_{ij}}{a^2_0} - D,
\label{ConfineP}
\\
V^{CS}_{ij} &=& \frac{\hbar^2 c^2 \kappa'}{m_i m_j c^4} \frac{e^{- \left( r_{ij} \right)^2 / \left( r_{0ij} \right)^2}}{(r_{0ij}) r_{ij}} \sigma_i \cdot \sigma_j\,.
\label{CSP}
\end{eqnarray}
Here, $r_{0 ij} = \left( \alpha + \beta m_{ij}  \right)^{-1}$, $\kappa' = \kappa_0 \left( 1 + \gamma m_{ij} \right)$ and $m_{ij}=\frac{m_i m_j}{m_i + m_j}$.

This model can explain the ground state hadron masses including light, charm, and bottom quarks\cite{Park:2018wjk,Noh:2021lqs}, and was also used to study possible compact exotic configurations\cite{Park:2018wjk,Noh:2021lqs,Park:2017jbn}.
Here, using a Gaussian wave function $\exp(-a^2 r^2)$ with $r$ being the length between the quark-antiquark pair, we fit the model parameters in the Hamiltonian to the ground state masses listed in Table~\ref{mesons1}.  The parameters are as follows.
$\kappa=90.0 \, \textrm{MeV fm}$, $a_0=0.0298664 \, \textrm{(MeV$^{-1}$fm)$^{1/2}$}$, $D=1084  \, \textrm{MeV}$, $ m_{u}=m_{d}=315 \, \textrm{MeV}$, $m_{s}=593 \, \textrm{MeV}$, $m_{c}=1883 \, \textrm{MeV}$, 	$\alpha = 1.0099 \, \textrm{fm$^{-1}$}$, $\beta = 0.0004314 \, \textrm{(MeV fm)$^{-1}$}$, $\gamma = 0.00108 \, \textrm{MeV$^{-1}$}$, $\kappa_0=220.144 \, \textrm{MeV}$.
The standard deviation of the masses obtained in Table~\ref{mesons1} is $
\sigma = ( \frac{1}{N-1} \sum_{i=1}^{N} \left( M^{Thr}_i - M^{Exp}_i \right)^2 )^{1/2} = 5.86~{\rm MeV},$ where  $M^{Thr}_i$ indicates the mass obtained by the model calculation and $M^{Exp}_i$ is the experimentally measured mass.

\begin{table}[ht]
\caption{Meson masses (Column 3) and the variational parameter $a$ (Column 4) from the present model calculation. $\sigma=5.86$ MeV.}

\centering

\begin{tabular}{cccc}
\hline
\hline	\multirow{2}{*}{Particle}	&	Experimental	&	Mass		&	\quad Variational\quad					\\
								&	Value (MeV)	&	(MeV)	&	\quad Parameter (${\rm fm}^{-2}$)\quad	\\
\hline  
$\eta_c$	&	2983.6	&	2996.9	&	\quad$a$ = 13.1\quad	\\
$J/\psi$	    &	3096.9	&	3089.6	&	\quad$a$ = 11.1\quad	\\
$D$         &	1864.8	&	1864.1	&	\quad$a$ = 4.5\quad	\\
$D^*$       &	2010.3	&	2010.7	&	\quad$a$ = 3.7\quad	\\
$\pi$		&	139.57	&	139.39	&	\quad$a$ = 4.6\quad	\\
$\rho$		&	775.11	&	775.49	&	\quad$a$ = 2.2\quad	\\
$K$		    &	493.68	&	494.62	&	\quad$a$ = 4.6\quad	\\
$K^*$		&	891.66	&	888.82	&	\quad$a$ = 2.8\quad	\\
\hline 
\hline
\label{mesons1}
\end{tabular}
\end{table}

\begin{table}[ht]
\caption{Baryon masses obtained with two-body forces only (bracketed values in Column 3) and after adding the three-body force (Colume3) from the model calculation in this work.   Column 4 shows the variational parameters $a_1$ and $a_2$. $\sigma = 25.90 (59.26)$ MeV with (without) three-body forces.}

\centering

\begin{tabular}{cccc}
\hline
\hline	\multirow{2}{*}{Particle}	&	Experimental 	&	Mass		&	\quad Variational\quad					\\
								& Value (MeV)	&	(MeV)	&	\quad Parameters (${\rm fm}^{-2}$)\quad	\\
\hline  
$\Lambda_{c}$	&	2286.5	&	2266.7 (2281.6)	&	\quad$a_1$ = 2.9, $a_2$ = 3.7\quad	\\
$\Sigma_{c}$	&	2452.9	&	2441.6 (2480.9)	&	\quad$a_1$ = 2.1, $a_2$ = 3.8\quad	\\
$\Lambda$		&	1115.7	&	1113.6 (1134.1)	&	\quad$a_1$ = 2.8, $a_2$ = 2.7\quad	\\
$\Sigma$		&	1192.6	&	1196.5 (1231.6)	&	\quad$a_1$ = 2.1, $a_2$ = 3.1\quad	\\
$\Xi_{cc}$		&	3621.2	&	3586.8 (3606.3)	&	\quad$a_1$ = 7.6, $a_2$ = 3.1\quad	\\
$\Sigma^*_{c}$&	2518.5	&	2522.9 (2567.7)	&	\quad$a_1$ = 2.0, $a_2$ = 3.4\quad	\\
$\Sigma^*$&	1383.7	&	1398.9 (1455.2)	&	\quad$a_1$ = 1.9, $a_2$ = 2.4\quad	\\

$p$				&	938.27	&	980.47 (1005.3)	&	\quad$a_1$ = 2.4, $a_2$ = 2.4\quad	\\
$\Delta$		&	1232	&	1272.1 (1346.8)	&	\quad$a_1$ = 1.8, $a_2$ = 1.8\quad	\\
\hline 
\hline
\label{baryons1}
\end{tabular}
\end{table}


{\it Baryon spectrum and quark three-body interaction:} We now use the same Hamiltonian and parameters to calculate the ground state baryon spectrum. The results are given in the bracket values in column 3 of Table~\ref{baryons1}. 
The two variational parameters $a_1$ and $a_2$ appearing in the table are the scaling factors of the Gaussian wave
function for the relative distance between the two identical quarks within the diquark and the relative distance
between the center of the diquark and the third quark, respectively.  
As can be seen in the Table, we find that the discrepancies for both the nucleon and Delta masses are large, and the overall 
$\sigma=59.26$ MeV is much larger than in the meson case.  This means that a quark model with two-body quark interactions successfully describing the meson spectrum fails to describe the baryon system.  In fact, it is known that the parameters have to be adjusted from mesons to baryons, even in a simple quark model \cite{Karliner:2014gca,Lee:2007tn}.  

Let us now introduce the quark three-body interaction given in Eq.~\eqref{eq:l3} with overall factors $A,B$ and $C$ multiplying each term.  Adding this to the two-body Hamiltonian given in Eq.~\eqref{Hamiltonian-two}, the total Hamiltonian becomes
\begin{eqnarray}
H_{Total} &=& H+ \sum^n_{i<j<k} \left( AL^{C-C}_{ijk} + BL^{S-S}_{ijk} + CL^{C-S}_{ijk} \right).
\label{Hamiltonian}
\end{eqnarray}
The calculated masses using the same two-body interactions but now including the three-body interactions are given in the third column of Table~\ref{baryons1}. 
As can be seen, the quality of the fit to the nucleon and Delta improves, and the overall $\sigma$ decreases to 25.90 MeV. The needed values for the three-body parameters are $A = -367.522 \, \textrm{MeV}^2$, $B = -2.85156 \times 10^{11} \, \textrm{MeV}^6$, $C = -7.68351 \times 10^6 \, \textrm{MeV}^4$. If one introduces a space dependence for the quark three-body interaction, the $\sigma$ becomes even smaller.  The results suggest that quark three-body forces exist and should be included for any quark configurations with quark and/or anti-quark numbers greater than three.

{\it Effects of quark three-body interaction on Exotics:} The newly introduced three-body interaction should play an important role in the stability of multiquark configuration. 
Table  \ref{tetraquarks-three-body}  shows the contribution of Eq.~\eqref{eq:l3} to the $T_{cc}$ and $\chi_{c1}(3872)$ tetraquarks.  Assuming the quantum numbers are $I(J^{P})=0(1^{+})$ for $T_{cc}$ and $I^G(J^{PC})=0^+(1^{++})$ for $\chi_{c1}(3872)$, the two independent color-spin bases for the former and latter can be conveniently chosen to be $(|\mathbf{3}_{12}\overline{\mathbf{3}}_{34}\rangle,|\overline{\mathbf{6}}_{12}\mathbf{6}_{34}\rangle)$ and 
$(|\mathbf{1}_{13}\mathbf{1}_{24}\rangle,|\mathbf{8}_{13}\mathbf{8}_{24}\rangle)$, respectively. Here, the subscripts $1,2,3,4$ represent  $\overline{q},\overline{q} ,c,c$ for  $T_{cc}$ and $\overline{c},\overline{q} ,c,q$ for $\chi_{c1}(3872)$, and the numbers in bold are the color multiplets of the quark pairs.  
It should be noted that the $f$-type interaction in $L^{S-S}$ shown in the last term of Eq.~\eqref{eq:l3} contributes to the (1,2) matrix element in Table III.  
This is the first case where the $f$-type interaction, which does not contribute to the masses of normal hadrons, affects the mass of  $T_{cc}$ with determined strength.

Solving for the ground state of $\chi_{c1}(3872)$ with two-body interactions only, one finds it to be a well-separated meson-meson state dominated by the $|\mathbf{1}_{13}\mathbf{1}_{24}\rangle$ color-spin state.  For the $T_{cc}$, the ground state is dominated by the $|\mathbf{3}_{12}\overline{\mathbf{3}}_{34}\rangle$ color state with the energy being slightly above or below the threshold, depending on the form of the $r$-dependence of the color-spin potential\cite{Noh:2021lqs,Noh:2023zoq}.  Using these ground states, we now calculate the expectation value of the quark three-body interaction. Table \ref{tetraquarks-three-body} shows the contributions of the three terms of the quark three-body potential using the same overall constants $A,B,C $ as determined from the baryon spectrum: this assumes that the spatial sizes are all similar.  For $T_{cc}$, the largest contribution comes from the term proportional to $d^{abc}$ in $L^{C-S}$, while for $\chi_{c1}(3872)$, it is from the first term of $L^{C-C}$ given in Eq.~\eqref{eq:l3}.  For both cases, one finds that the sums of the quark three-body interactions are repulsive of more than 10 MeV, 
which places the masses of compact configurations far above the thresholds and pushes the quarks into separate meson states.

In this work, we have derived the quark three-body interactions and shown they are essential for systematically applying quark models from a two-body system (meson) to a three-body system (baryon). We have further demonstrated that such quark three-body interactions provide a non-trivial amount of repulsion in a compact configuration for both the $T_{cc}$ and 
$\chi_{c1}(3872)$, strongly suggesting that these will fall apart and become molecular-type meson systems bound by pion exchange type of interaction with large sizes\cite{Yun:2022evm} where the quark three-body force will be small. Hence, one can conclude that such quark three-body interactions are crucial in the study of compact quark configurations with quark and anti-quark numbers larger than three.

\begin{widetext}

\begin{table}[ht]
\caption{Quark three-body potentials for $T_{cc}$ and $\chi_{c1}(3872)$. The color basis set for two states are $(|\mathbf{3}_{12}\overline{\mathbf{3}}_{34}\rangle,|\overline{\mathbf{6}}_{12}\mathbf{6}_{34}\rangle)$ and $(|\mathbf{1}_{13}\mathbf{1}_{24}\rangle,|\mathbf{8}_{13}\mathbf{8}_{24}\rangle)$, respectively.}

\centering

\begin{tabular}{ccc}
\hline
\hline		&	$T_{cc}$	&	$\chi_{c1}(3872)$			\\
\hline  
$\sum L^{C-C}_{ijk}$	&	$\left(
    \begin{array}{cc}
       \frac{32}{9m_u}+\frac{32}{9m_c}  & 0  \\
       0  & -\frac{208}{9m_u}-\frac{208}{9m_c}
    \end{array} \right)$	&	$\left(
    \begin{array}{cc}
       -\frac{128}{9m_u}-\frac{128}{9m_c}  & -\frac{80\sqrt{2}}{9m_u} -\frac{80\sqrt{2}}{9m_c}  \\
       -\frac{80\sqrt{2}}{9m_u} -\frac{80\sqrt{2}}{9m_c}  & -\frac{16}{3m_u}-\frac{16}{3m_c}
    \end{array} \right)$	\\
$\sum L^{S-S}_{ijk}$	&	\tiny{$\left(
    \begin{array}{cc}
       -\frac{224}{9m_u^3 mc^2}+\frac{224}{3m_u^2 m_c^3}  & -\frac{224\sqrt{2}}{3m_u^4 m_c} +\frac{32\sqrt{2}}{m_u^3 m_c^2} - \frac{32\sqrt{2}}{m_u^2 m_c^3} + \frac{160\sqrt{2}}{3m_u m_c^4} \\
       -\frac{224\sqrt{2}}{3m_u^4 m_c} +\frac{32\sqrt{2}}{m_u^3 m_c^2} - \frac{32\sqrt{2}}{m_u^2 m_c^3} + \frac{160\sqrt{2}}{3m_u m_c^4}  & -\frac{112}{3m_u^3 m_c^2}+\frac{112}{9m_u^2 m_c^3}
    \end{array} \right)$}	&	\tiny{$\left(
    \begin{array}{cc}
       -\frac{128}{9m_u^3 mc^2}-\frac{128}{9m_u^2 m_c^3}  & -\frac{80\sqrt{2}}{9m_u^3 m_c^2} -\frac{80\sqrt{2}}{9m_u^2 m_c^3}  \\
       -\frac{80\sqrt{2}}{9m_u^3 m_c^2} -\frac{80\sqrt{2}}{9m_u^2 m_c^3}  & \frac{32}{3m_u^3 m_c^2}+\frac{32}{3m_u^2 m_c^3}+\frac{16}{m_u m_c^4}+\frac{16}{m_u^4 m_c}
    \end{array} \right)$}	\\
$\sum L^{C-S}_{ijk}$ & $\left(
    \begin{array}{cc}
       -\frac{256}{3m_u^3}+\frac{256}{9m_c^3}  & \frac{32\sqrt{2}}{3m_u^2 m_c} +\frac{32\sqrt{2}}{3m_u m_c^2}  \\
       \frac{32\sqrt{2}}{3m_u^2 m_c} +\frac{32\sqrt{2}}{3m_u m_c^2}  & -\frac{320}{9m_u^3}+\frac{320}{3m_c^3}
    \end{array} \right)$ & $\left(
    \begin{array}{cc}
       \frac{256}{9m_u^2 m_c}+\frac{256}{9m_u m_c^2}  & \frac{160\sqrt{2}}{9m_u^2 m_c} +\frac{160\sqrt{2}}{9m_u m_c^2}  \\
       \frac{160\sqrt{2}}{9m_u^2 m_c} +\frac{160\sqrt{2}}{9m_u m_c^2} & -\frac{16}{m_u^3}-\frac{16}{3m_u^2 m_c} -\frac{16}{3m_u m_c^2}-\frac{16}{m_c^3}
    \end{array} \right)$ \\
\hline 
\hline
\label{tetraquarks-three-body}
\end{tabular}
\end{table}

\begin{table}[ht]
\caption{Masses of $T_{cc}$ and $\chi_{c1}(3872)$ obtained without (bracket values in Column 3) and with (Column 3) three-body forces.  Columns 4-6 show the contribution of the three-body force in MeV. Column 7 shows the variational parameters $a_1$, $a_2$ and $a_3$ defined in Refs.~\cite{Noh:2021lqs,Noh:2023zoq}. }

\centering

\begin{tabular}{ccccccc}
\hline
\hline	Particle	&	Measured mass (MeV)	&	Mass (MeV)	&  $\sum_{i<j<k}L^{C-C}_{ijk}$  & $\sum_{i<j<k}L^{S-S}_{ijk}$  &  $\sum_{i<j<k}L^{C-S}_{ijk}$ &	Variational parameters (${\rm fm}^{-2}$) 	\\
\hline  
$T_{cc}$	&	3875	&	3971.70 (3955.18)	& -3.86792    & -0.458326  & 20.8477  &		\quad$a_1$ = 2.8, $a_2$ = 7.3, $a_3$ = 2.7\quad	\\
$\chi_{c1}(3872)$	&	3872	&	3884.25 (3866.20)	& 19.3694   & 0.0427164  & -1.36541  &		\quad$a_1$ = 11.1, $a_2$ = 2.2, $a_3$ = 0.01\quad	\\
\hline 
\hline
\label{tetraquarks}
\end{tabular}
\end{table}

\end{widetext}

{\it Acknowledgements:}  
This work was supported by the Korea National Research Foundation under Grant No. 2023R1A2C300302311 and 2023K2A9A1A0609492411.
S. Noh and S. Cho was supported by the
National Research Foundation of Korea (NRF) grant funded by the Korea government (MSIT) (No. RS-2023-
00280831).

\end{document}